# The Emergence of Anisotropic Superconductivity in the Nodal-line Semi-metal TlTaSe$_2$

Mukhtar Lawan Adam[1,2,3]*, Ibrahim Buba Garba[4,5], Sulaiman Muhammad Gana[3], Bala Ismail Adamu[6,7], Abba Alhaji Bala[6], Abdulsalam Aji Suleiman[8†], Ahmad Hamisu[9], Tijjani Hassan Darma[3], Auwal Musa[3] and Abdulkadir S. Gidado[3]

[1] Information Materials and Intelligent Sensing Laboratory of Anhui Province, Key Laboratory of Structure and Functional Regulation of Hybrid Materials of Ministry of Education, Institutes of Physical Science and Information Technology, Anhui University, Hefei 230601, People's Republic of China

[2] Materials Interfaces Center, Shenzhen Institute of Advanced Technology, Chinese Academy of Sciences, Shenzhen 518055, Guangdong, PR China

[3] Physics Department, Bayero University, Kano 700231, Nigeria

[4] Sorbonne Université, CNRS, IMPMC, 4 Place Jussieu, Paris 75252, France

[5] Department of Physics, Federal University Gashua, Gashua 671106, Nigeria

[6] Physics Department, Federal University Dutse, Jigawa 720101, Nigeria

[7] School of Emergent Soft Mater, South China University of Technology, Guangzhou International Campus, Guangzhou 510640, People's Republic of China

[8] Institute of Materials Science and Nanotechnology, Bilkent University UNAM, Ankara 06800, Turkey

[9] Physics Department, King Abdulaziz University, Jedda 21589, Saudi Arabia

Mukhtar Lawan Adam and Ibrahim Buba Garba contributed equally

Corresponding authors: * mladam.phy@buk.edu.ng; † abdulsalam@unam.bilkent.edu.tr

**Abstract**

TlTaSe$_2$ is a non-centrosymmetric quasi-2D crystal semi-metal hosting nodal-line topological features protected by mirror-reflection symmetry. Here, we investigated the superconducting properties of TlTaSe$_2$ using the first-principles anisotropic Migdal-Eliashberg theory. The Fermi surface hosts well gapped multiband features contributed by the Ta 5*d* and Tl 6*p* orbitals. Moreso, anisotropic superconducting gaps were found to exist at 2.15 and 4.5 meV around the in-plane orbitals, coupling effectively with the in-plane phonons of the Ta and Tl atoms. Using the Allen-Dynes-modified McMillan formula, we found a superconducting transition temperature of 6.67 K, accompanied by a robust electron-phonon coupling constant λ of 0.970. This investigation provides valuable insights into the mechanisms underlying anisotropic superconductivity in TlTaSe$_2$.

**Keywords**: TlTaSe$_2$, Anisotropic superconductivity, Quantum materials, Nodal-line Semimetal

**Introduction**

The interplay of anisotropic superconductivity and nontrivial topological states can manifest emergent quantum electronic behaviors in materials. [1,2] Anisotropy in superconductors manifests as discernible variations in superconducting characteristics at the Fermi surface, primarily attributable to the intrinsic anisotropy found within their electronic states and lattice vibrational modes. This anisotropy exerts a significant influence on a spectrum of superconducting properties, encompassing the critical magnetic field, penetration depth, and the magnitude of the energy gap linked to Cooper pairs, ultimately governing the material's behavior in diverse magnetic and physical environments. [1–3] However, the underlying mechanisms driving superconductivity in materials are electron-phonon interactions, which are generally considered isotropic in metals. [4,5] Recent studies suggest that electron-phonon (el-ph) coupling can be anisotropic because of phonon anharmonicity [6], islands of isolated electronic states, and van Hove singularities around the Fermi surface. [5]

Intercalated layered transition metal dichalcogenide compounds have potential anisotropic superconducting properties. [7–10] Previous studies demonstrated that TlTaSe$_2$ is a topological nodal-line semi-metal with robust nontrivial topological surface states. [11] Exploring materials with anisotropic superconductivity and robust topological features is crucial for advancing the fundamental understanding of topological superconductivity and potentially discovering novel applications in superconducting technologies. [2,3]

Herein, we investigate the superconducting properties and electronic structures of TlTaSe$_2$. Electronic band structure analysis reveals the material's topological nodal line semi-metal nature, with a distinct drum-like feature in the bands indicating its properties. Surface state calculations employing the Wannier method revealed robust topological surface states on the 001 crystallographic surfaces. From the study of the el-ph interactions, strong anisotropic el-ph coupling exists at the multiple states around the Fermi surface. The strength of this interaction, which governs the superconducting transition temperature and gap structure, was analyzed through phonon dispersion and electron-phonon coupling strength. This investigation provides valuable insights into the mechanisms underlying anisotropic superconductivity in TlTaSe$_2$.

**Computational methods**

As implemented in Quantum Espresso, the first-principles calculations were performed within the density functional theory using norm-conserving pseudopotential. [12] [13] A kinetic energy cutoff of 60 Ry was used in the self-consistent calculations. We sampled the Brillouin zone with a gamma-centered k-mesh of 15 × 15 × 7. We also used the wannier90 code [14] to obtain the materials' maximally localized functions; were the surface Green's function [15] was used, as implemented in the WannierTools [14] package, to calculate the surface state dispersion of the materials. The dynamical matrices and linear variation of the self-consistent potential were calculated within the density-functional perturbation theory [16] on the irreducible set of regular $3^3$ **q**-point meshes. The electronic wave functions for the Wannier interpolation within the EPW were calculated on uniform and gamma-centered **k**-point meshes of size $6^3$. To solve the Eliashberg equations, we evaluate the electron energies, phonon frequencies, and electron-phonon matrix elements in a fine grid using the method of Giustino et al. [17]. The fine grids contain $12^3$ **q**-points and $24^3$ **k**-points uniform gamma-centered grids. The Eliashberg function α$^2$F(ω) and cumulative contribution to the electron-phonon coupling strength λ(ω) were obtained from [18]

$$\alpha^2 F(\omega) = \frac{1}{2\pi N_{Ef}} \sum_{qv} \frac{\lambda_{qv}}{\omega_{qv}} \delta(\omega - \omega_{qv}) \qquad (1)$$

$$\lambda(\omega) = 2\int_0^\omega \frac{\alpha^2 F(\omega')}{\omega'} d\omega' \qquad (2)$$

Where $\lambda_{qv}$ is phonon wave vector **q** with v mode resolved electron-phonon coupling constant and $N_{Ef}$ is the density of states at the Fermi level.

We used the Allen-Dynes formula, equation (3) to estimate the Tc,

$$Tc = \frac{\omega_{log}}{1.2} \exp\left(-\frac{1.04(1+\lambda)}{\lambda - \mu^*(1 + 0.62\lambda)}\right), \qquad (3)$$

where $\omega_{log}$ is the logarithmic averaged phonon frequency

$$\omega_{log} = \exp\left(\frac{2}{\lambda}\int \frac{\alpha^2 F(\omega) \log(\omega)}{\omega} d\omega\right),$$

**Results and discussions**

**Crystal structure TlTaSe$_2$**

TlTaSe$_2$ has a non-centrosymmetric crystal structure that belongs to the P-6m2 (187) space group. [11] A unit cell hosts one atom each of Tl, Ta, and two atoms of Se. The Tl layer is intercalated between layers of two tantalum dichalcogenide layers, with Tl atoms aligned with Ta atoms in the vertical direction, as shown in **Figure 1(a)**. The mirror plane in this crystal is symmetric along the Tl and Ta planes, which plays a significant role in nodal line protection in this type of structure. [19]

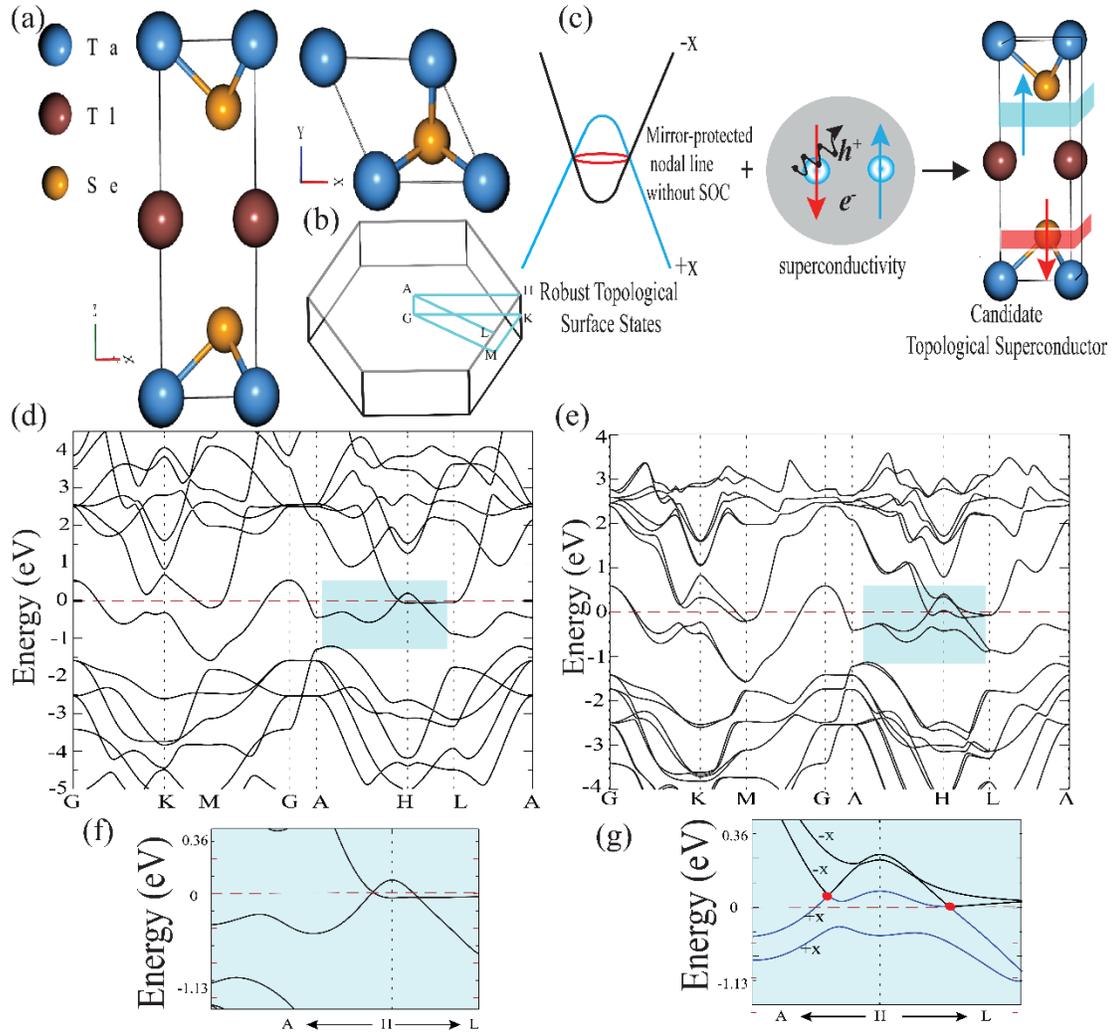

**FIG. 1. (a)** Side and top views of TlTaSe$_2$ crystal structure. **(b)** Brillouin zone with high-symmetry points. **(c)** Schematic representation of the work, highlighting the coexistence of robust topological surface states and superconductivity in material, making it a candidate for the search for topological superconductivity. Calculated band structure **(d)** without spin-orbit coupling inclusion **(e)** with spin-orbit coupling. Close-up view of the bands crossing **(f)** without spin-orbit coupling **(g)** with spin-orbit coupling.

**Electronic structure TlTaSe$_2$**

The electronic band structures of TlTaSe$_2$, both in the presence and absence of spin-orbit coupling (SOC), are shown in **Figures 1(d)** and **1(e)**, respectively. The inclusion of SOC in the band structure calculations (**Figure 1(e)**) was due to the absence of inversion symmetry in the crystal structure. The SOC splits the bands into two bands, each resulting in four. The four resultant bands can be classified with mirror eigenvalues of ±x, as shown in **Figure 1(g)**. The two-hole-like bands have mirror eigenvalues of -x, while the two electrons-like bands have +x mirror eigenvalues. Moreover, there is band inversion around the **H** high-symmetry point. Combined with time-reversal symmetry, this band inversion can result in a protected nodal line protected by non-symmorphic space group symmetries that survive even in the presence of SOC. [20]

Focusing on the vicinity of the **H** high-symmetry point, **Figures 1(f)** and **1(g)** provide an enlarged view of the band crossing. Notably, in **Figure 1(d)**, two discernible bands are the predominant features near the Fermi level. This observation is further elucidated in **Figure 2**, where the orbital characteristics surrounding the H high-symmetry points demonstrate substantial hybridization between a hole band originating primarily from the 5$d$ orbitals of the Ta atoms and an electron band attributed to the 6$p$ orbitals of the Tl atoms. Hybridization of these orbital states can have significant implications for the material's behavior. **Figure 2(d)** shows the Fermi surface of bulk TlTaSe$_2$ with distinct islands of electronic and hole sheets around the **G** and **M** high-symmetry points.

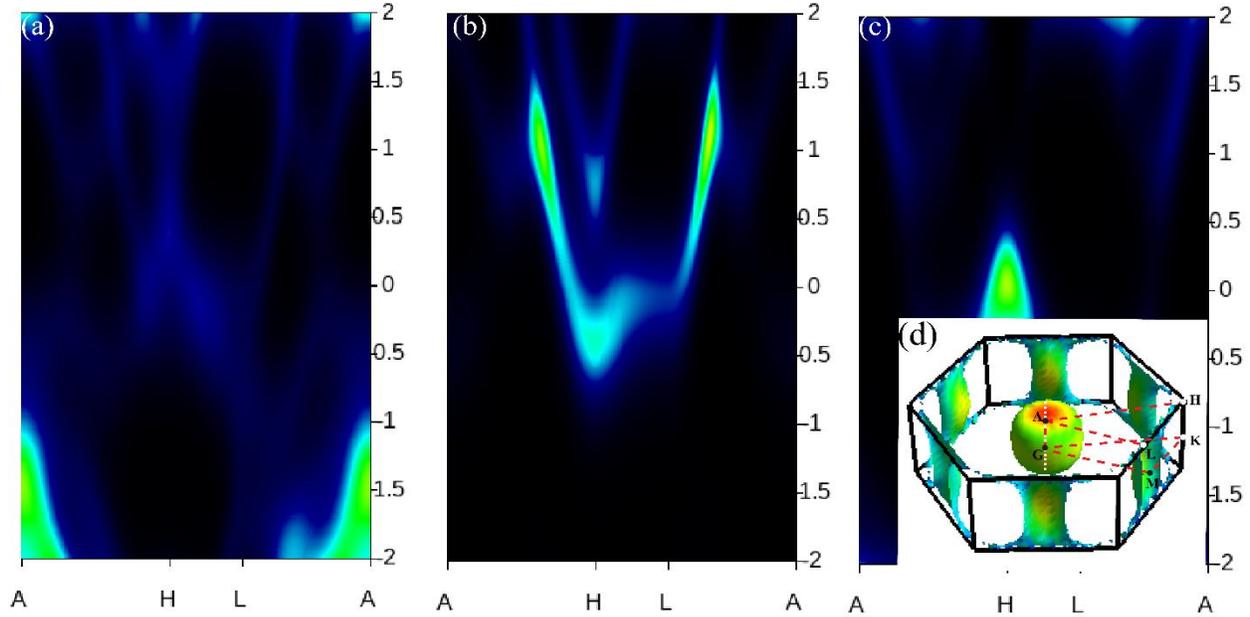

**FIG. 2**. High symmetry points projected density of states: **(a)** Total density of states. (b) The density of state contribution by Tl $p$ atomic orbitals. (c) The density of state contribution by Ta $d$ atomic orbitals. (d) The Fermi surface of bulk TlTaSe$_2$ shows two distinct Fermi surface sheets around the **A** and **G** (S$_1$) and **L** and **M** (S$_2$) high-symmetry points.

Furthermore, distinctive characteristics commonly associated with topological nodal line semi-metals include resilient topological surface states and Fermi arcs. Therefore, we explored the topological features of TlTaSe$_2$. Our investigation delves into calculating the surface states and Fermi arcs along the **A – H – L** direction on the (001) surface. To carry out this analysis, we employed a slab model and utilized the recursive Green function method, implemented within a Wannier-based tight-binding Hamiltonian framework. The resultant surface band structures pertain to two feasible surface terminations within the TlTaSe$_2$ system: the Se (001)-and Tl (010)-terminated surfaces. **Figure 3** shows the topological features of the Se (001)-terminated surface. The band structures obtained for the terminated surface reveal intriguing patterns, mainly characterized by the emergence of drum-like surface states above the Fermi level around the **k**- high symmetry point. As observed in **Figures 3(a)** and **(c)**, the 001 terminated surface has robust surface states (SS). We created a model slab with 50 TlTaSe2 unit cells to ascertain these states further, as shown in **Figure 3(b)**. The bulk states and topological surface states can be distinctively observed and visualized experimentally through ARPES experiments. In **Figure 3(d)**, the Fermi surface presents islands of non-nested electronic sheets, which can give rise to possible anisotropy in TlTaSe$_2$ electronic properties.

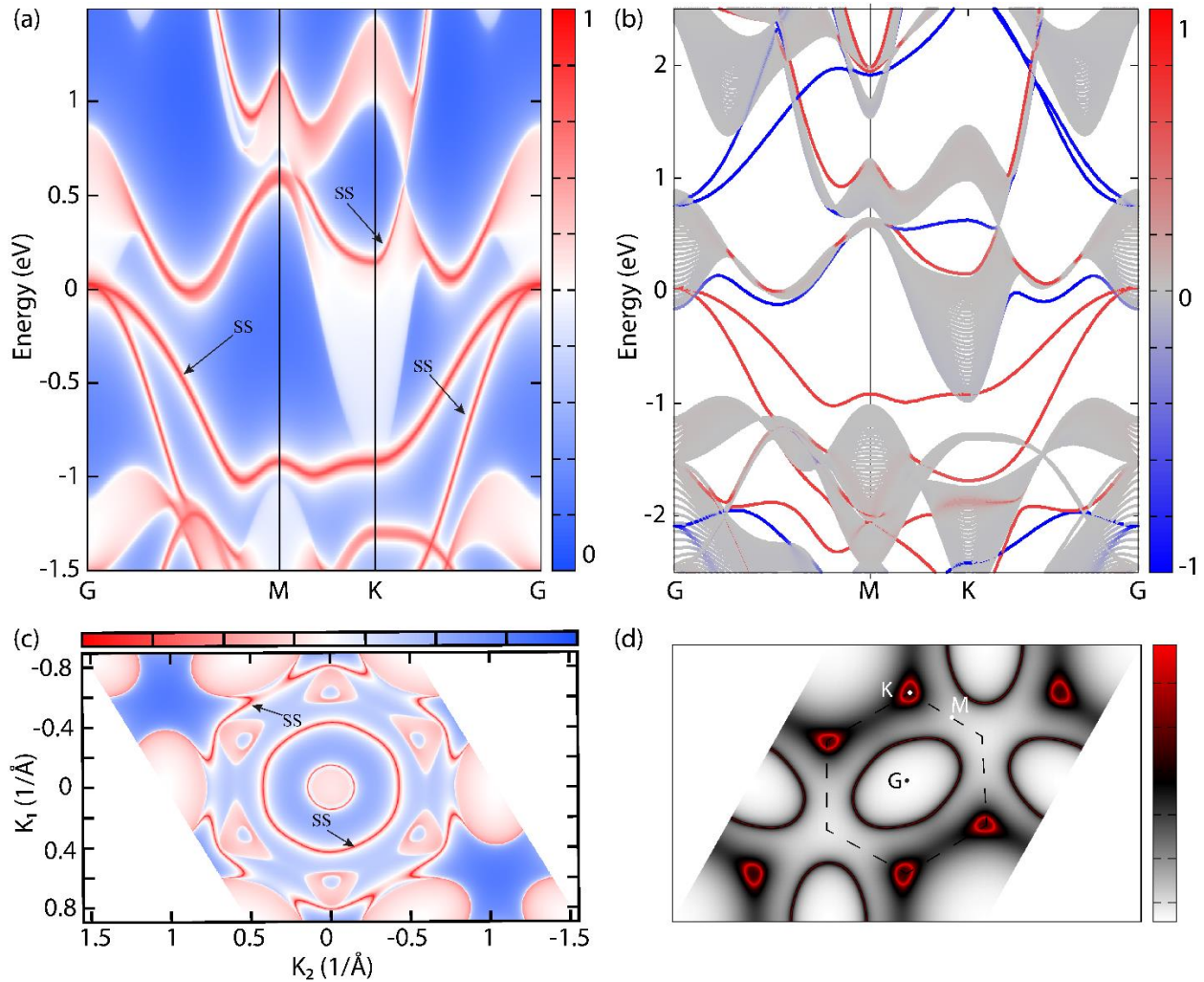

**FIG. 3. (a)** Bulk bands projected on the (001) surface of TlTaSe$_2$. The nontrivial topological surface states (SS) are indicated with arrows. **(b)** Surface weighted energy bands of 50-unit-cell slab (001) surface of a slab model of TlTaSe$_2$. The blue and red nontrivial topological states are contributions from the Se and Tl surfaces, respectively. **(c)** Isoenergy-band contours at $E = E_f$ around the **K** high symmetry point, showing nontrivial surface states (SS). **(d)** The surface Fermi states of the 001-projected TlTaSe$_2$ slab at **k**z = 0.

**Superconducting properties of TlTaSe$_2$**

The electronic states residing at the Fermi level have substantial significance in dictating the superconducting behavior observed in materials. Anisotropic superconductivity is a distinctive phenomenon in various materials that can emerge due to Fermi surface sheet anisotropy. [21] **Figure 4(a)** offers insight into the calculated phonon dispersion of TlTaSe$_2$, accompanied by the corresponding phonon density of states. The isotropic Eliashberg spectral function $\alpha_2F(\omega)$, depicted in **Figure 4(b)**, and the el-ph coupling strength $\lambda(\omega)$ reveal a notable feature: a distinct gap centered around 20 meV that cleaves the spectrum into two segments, as evidenced in **Figure 4(c)**. This division indicates the presence of distinct vibrational modes, each of which contributes to the material's response.

Delving into the nature of these vibrational modes, those with frequencies lower than 20 meV arise predominantly from the in-plane and out-of-plane oscillations of the Tl and Ta atoms. In contrast, the higher frequency modes, surpassing 20 meV, are chiefly attributed to vibrations involving Se atoms. This distinction in vibrational frequencies can be ascribed to the considerably larger atomic masses of Tl and Ta compared to those of Se atoms. The calculated isotropic Eliashberg spectral function and e-ph coupling strength, as shown in **Figure 4(b)**, revealed the prevalence of robust el-ph coupling interactions. These interactions were particularly pronounced in the low-frequency modes below 20 meV. Remarkably, these interactions contribute to over 85% of the total electron-phonon coupling strength, underscoring their dominance in dictating the electronic and vibrational behavior of the material.

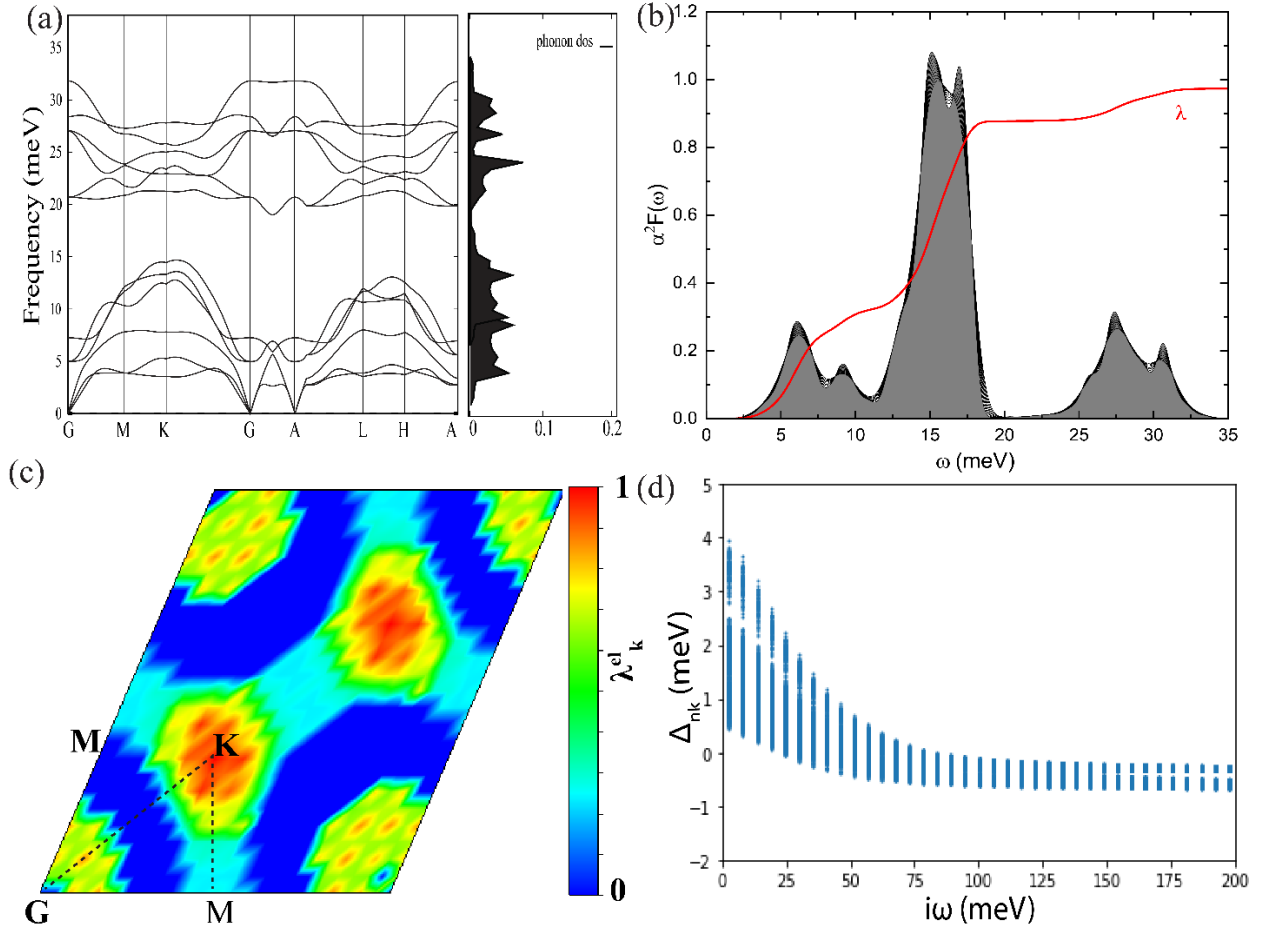

**FIG. 4**. Calculated electron-phonon coupling and superconducting properties of TlTaSe$_2$. **(a)** Phonon dispersion and calculated partial phonon density of states (DOS) of TlTaSe$_2$. **(b)** Isotropic Eliashberg spectral function $\alpha^2F(\omega)$ and integrated electron-phonon coupling strength $\lambda(\omega)$ (red line). **(c)** Momentum-resolved electron-phonon coupling strength $\lambda^{el}_{\mathbf{k}}$ on the Fermi surface. **(d)** Calculated energy-dependent superconducting gap of TlTaSe$_2$ at T = 10 K. The gap was obtained by solving the anisotropic Eliashberg equation with $\mu_c^* = 0.16$.

Furthermore, continuing our exploration, we estimated the critical superconducting transition temperature ($T_c$) using the Allen-Dynes-modified McMillan formula [22] with a chosen Coulomb pseudopotential $\mu^* = 0.16$. The outcome reveals a superconducting transition temperature of $T_c =$

6.67 K, accompanied by a robust el-ph coupling constant λ = 0.970 and a logarithmic phonon frequency $\omega_{log}$ = 12.161 meV. These parameters collectively provide insights into the propensity of the material for superconductivity and the strength of its electron-phonon interactions. By employing the Anisotropic Migdal-Eliashberg theory using Wannier functions [23], as shown in **Figure 4(c)**, we depict the momentum-resolved electron-phonon coupling, allowing us to gain a nuanced understanding of the electron-phonon interaction strengths associated with the Fermi surface sheets present in TlTaSe$_2$. Notably, the electron-phonon coupling strength near the **K** high-symmetry point surpassed that of the sheet around the **G** point. This observation underscores the anisotropic distribution of the electron-phonon interactions across the Fermi surface.

Consequently, we studied the superconducting gap inherent to TlTaSe$_2$. As illustrated in **Figure 4(d)**, the energy gaps manifest as two discernible entities, revealing the anisotropic characteristics of the superconducting state. These energy gaps can be attributed to an average energy of approximately 2.15 meV for the $\Delta_1$ Fermi-surface sheets and around 4.50 meV for $\Delta_2$, as elucidated in **Figures 5(a)** and **(b)**.

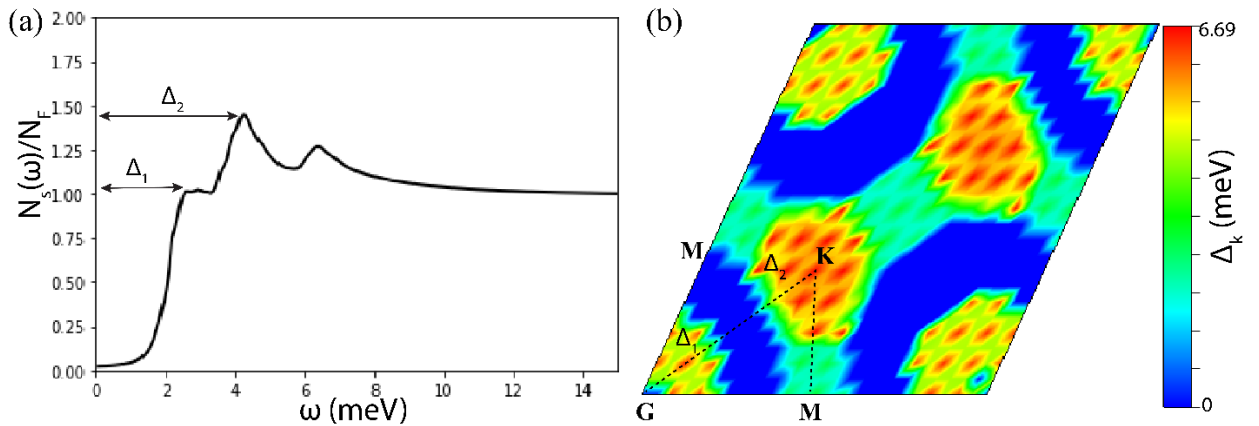

**FIG. 5. (a)** Calculated normalized quasiparticle density of states of TlTaSe$_2$ as a function of frequency at 10 K. **(b)** Momentum-resolved superconducting gap $\Delta_\mathbf{k}$ on different sheets of the Fermi surface of TlTaSe$_2$.

The anisotropic superconductivity observed in TlTaSe$_2$ arises from the complex interplay between the distinct Fermi surfaces and phonon contributions of its constituent atoms. This interrelationship gives rise to intriguing and tunable superconducting behavior that is pivotal for understanding the material's unique properties. The phonon contributions originating from the atomic vibrations of the Tl, Ta, and Se atoms play a crucial role in determining the anisotropic nature of the superconducting state. [24,25] These atomic vibrations, characterized by different frequencies, reflect the distinct masses of the atoms, with heavier Tl and Ta atoms primarily contributing to lower-frequency vibrations and lighter Se atoms dominating the higher-frequency range. This imparts pronounced anisotropy to the superconducting properties of TlTaSe$_2$.

**Conclusions**

In summary, the anisotropic superconductivity observed in TlTaSe$_2$ can be viewed as an intricate interplay between the distinct Fermi surface sheets and the phonon contributions of its constituent atoms. This relationship underscores the material's versatile and tunable superconducting properties, making it an intriguing platform for exploring anisotropic superconductivity and its potential applications in emerging technologies. The intricate interplay between electronic structure, phonon dynamics, and Fermi surface topology deepens our understanding of superconducting mechanisms and opens doors to engineering novel materials with tailored anisotropic properties for advanced electronic and quantum devices.

**Acknowledgements**

M.L.A. acknowledges support from Bayero University Kano, Anhui University and China Postdoctoral foundation for the postdoctoral. The National Synchrotron Radiation Laboratory of China, Hefei, People's Republic of China, for access to their supercomputing facilities. Finally, I.B.G. acknowledges the PTDF Fund.

## Author Contributions

M.L.A. conceived the idea and carried out the experiments and calculations. M.L.A., T.H.D. and A.A.S. supervised the project. M.L.A., I.B.G., and S.M.G. analyzed the results and prepared the draft copy. All the authors participated in discussing the results, analysis, and revision of the final manuscript.

## Declaration of interests

The authors declare no competing interests, either financial or otherwise.

Data